# Structural social capital and health in Italy [1]

Damiano Fiorillo[2] and Fabio Sabatini [3] [4]

**Abstract**

This paper presents the first empirical assessment of the causal relationship between social capital and health in Italy. The analysis draws on the 2000 wave of the Multipurpose Survey on Household conducted by the Italian Institute of Statistics on a representative sample of the population ($n$ = 46,868). Our measure of social capital is the frequency of meetings with friends. Based on IV and bivariate probit estimates, we find that individuals who meet friends every day or at least two times a week are approximately 11% to 16% more likely to report good health.

[1] We are indebted to two anonymous referees whose comments allowed a substantial improvement of the paper. We thank Luigi Aldieri, Francesco Drago, Elena Fumagalli and Alessandra Gualtieri for precious comments and suggestions. Needless to say, usual caveats apply.

[2] Department of Business and Economics, University of Napoli Parthenope, and Health, Econometrics and Data Group, University of York. Email: damiano.fiorillo@uniparthenope.it.

[3] Department of Economics and Law, Sapienza University of Rome, Italy, and Laboratory for Comparative Social Research, National Research University Higher School of Economics, Russia. Email: fabio.sabatini@uniroma1.it.

[4] Corresponding author. Postal address: Facoltà di Economia, Dipartimento di Economia e diritto, via del Castro Laurenziano 9, 00161, Roma, Italy. Phone and fax: +39 06 49766949.

# 1. Introduction

The claim that social capital plays a role in determining actual and perceived health is commonly accepted in public health studies (Kawachi et al. 1997; Kawachi et al. 1999; Kim et al. 2006; Kim et al., 2011) and has recently attracted the attention of economists and economics journals (Brown et al. 2006; Petrou and Kupek 2008; Scheffler and Brown 2008; D'Hombres et al., 2010). Two critical issues have emerged from previous research on the topic.

First, social capital is a very multidimensional phenomenon and there is no univocal evidence on which of its dimensions is good for health. The relationship between the multiple facets of social capital and health is context-dependent and varies according to a number of individual, social, and institutional features.

Second, even if many studies identify social capital as a significant predictor of individual health, there are reasons to suspect this result to be due to a spurious correlation. It seems reasonable to assume the existence of reverse causality: unhealthy people may face obstacles to social interaction, while healthy people may be more inclined to certain relational activities such as, for example, doing sports with others.

The present paper contributes to the literature by carrying out the first assessment of the causal relationship between social capital and individual health in Italy. Similar research has been undertaken in North America (see for example Folland 2007), Latin America (Ronconi et al. 2010), and Eastern Europe (D'Hombres et al. 2010) but, to the best of our knowledge, they have never been performed in Mediterranean countries.

Probit estimates show that, in addition to civil status, age, education, income and work status, individual structural social capital, as measured by the frequency of meetings with friends, is strongly and positively correlated with self-perceived health. However, since the habit of meeting friends may be endogenously determined, we follow some promising previous studies (see for example D'Hombres et al. 2010) and instrument this variable with Mass attendance and meetings with friends at the community level.

Instrumental variables regressions show that the habit of meeting friends is a strong predictor of perceived good health both with two-stages probit estimators and bivariate probit estimators.

The reminder of the paper is structured as follows. The next Section reviews the literature on social interactions and health and briefly presents our hypotheses. We then describe methodology and data. Section four describes and discusses empirical results. Concluding remarks and a brief discussion of policy implications close the paper.

## 2. Related literature

Over the past 20 years, the literature has extensively analyzed the impact of social interactions on individual health. Various aspects of the relational sphere of individual lives have been addressed, from relationships with family and friends to membership of various kinds of associations and community cohesion, often grouped together under the common label of social capital. After Putnam's seminal work (1993; 1995), social capital is usually referred to as "features of social organization such as networks, norms, and social trust that facilitate coordination and cooperation for mutual benefit." (Putnam 1995, 65). In Putnam's view social capital is a shared resource that serves to facilitate collective action. In this paper we focus on the micro dimension of the concept and refer to the definitions of Bourdieu (1986) and Coleman (1988) according to whom social capital is an individual resource that agents can access through social relationship.

Bourdieu (1980) argues that actors may use social relations as means to increase their ability to advance personal interests and improve well-being. In this context, social capital is "the sum of the resources, actual or virtual, that accrue to an individual or group by virtue of possessing a durable network of more or less institutionalized relationships of mutual acquaintance and recognition" (Bourdieu and Wacquant, 1992: 119, expanded from Bourdieu, 1980: 2).

Coleman (1988) considers social capital as a resource that, while being inherent in the structure of relations between actors, basically serves to "facilitate certain actions of actors, whether persons or corporate actors, within the structure" (p. 98). In his work, Coleman (1988; 1990) basically refers to the concept as "a resource for persons": each individual holds definite endowments of social capital and aims to use them to pursue her particular interests.

These perspectives describe social capital as a multidimensional concept, whose more tangible – and measurable – expression seems to consist of networks of human relations. Putnam (1995) argued that a research priority is to clarify the dimensions of social capital. We largely adopt the distinction between the structural and cognitive dimensions of social capital drawn by Uphoff (1999) in regard to the structural dimension. Structural social capital concerns individuals' behaviours and mainly takes the form of networks, which can be observed and measured through surveys. Cognitive social capital derives from individuals' perceptions resulting in norms, values and beliefs that contribute to cooperation.

More specifically, we focus on the structural dimension of social capital and measure it at the individual level as the frequency of meetings with friends, which are a commonly accepted proxy of social capital that has been used in the public health literature (see for example Kim and Kawachi, 2006; Baron Epel et al., 2008; Berry and Welsh, 2010).

We believe that this specific dimension of social capital can explain variations in individual health better than the related concepts of social trust and membership in associations[5]. It is worth noting that interactions with friends are different from the concept of social support. Meetings with friends, in fact, do not necessarily provide individuals with social (instrumental, emotional, and informational) support. Whether friendships actually have this function is likely to depend on a number of social and cultural circumstances. For example, friends may be a source of exposure to unhealthy behaviours. Interactions with friends may also cause sentiments of envy, low self-esteem, and relational anxiety (Steinfield et al., 2008). Or, more simply, friendships may be incapable of providing emotive support and/or access to unknown information. Using interactions with friends as a measure of social support, on the other hand, may lead to a tautology. In fact, a research identifying social support with meetings with friends would necessarily find that these meetings by definition entail support, which is a circumstance that, on the contrary, varies with circumstances and needs to be empirically tested, which is the objective of our empirical analysis.

---

[5] It is worth noting that several studies do not find any correlation between social trust and perceived health (Kennelly et al. 2003; Carlson 2004; Sabatini, 2014). Some studies find that the association between social trust and various measures of health is not robust to different specifications (Kim et al. 2006; Folland 2007; Baron-Epel et al. 2008; Mansyur et al. 2008). While results from the empirical literature seem to be somewhat conflicting, to date we lack a theoretical explanation of the causal mechanism possibly connecting social trust with health, which induces the reasonable suspect that studies ascertaining a significant relationship between trust and health actually found a correlation.

We argue that social interactions with friends may improve health through four channels:

1) *Transmission of health information*. Networks of relationships are a place to share past experiences on diseases, doctors, health facilities and therapies. This channel of information fosters matching procedures (in the sense that patients spend less time finding the appropriate doctor), lowers the cost of health information, speeds up the diffusion of knowledge of health innovation and eliminates mistaken perceptions on the role of healthcare, discouraging patients from undertaking inappropriate treatments.

2) *Mutual assistance mechanisms*. In case of sickness, the support of family and friends plays a fundamental role in ensuring access to healthcare services and facilities, for example through financial assistance, transportation services and help in dealing with doctors. Social contacts may foster individual access to services even when public protection schemes are designed to provide universal coverage (van Doorslaer et al. 2004). For example, empirical evidence on the Italian National Health System (NHS) – which theoretically covers all citizens on equal terms – suggests that the wealthy are more likely to be admitted to hospital than the poor (Masseria and Giannoni 2010). With reference to Italy, Atella et al. (2004) find that individuals who might be considered vulnerable from a societal perspective – i.e. the sick, the elderly, women and those with low incomes – are less likely to seek care from specialists and more likely to seek care from general practitioners. Since, in the Italian NHS, services are accessible by all citizens on a universal basis, health inequalities may also be related to people's ability to acquire suitable information and to find the right contacts in the right places, which in turn is influenced by the extension of one's social network.

3) *Promotion of healthy behaviours*. Social interactions may foster the development of social norms that support health-promoting behaviours, such as prevention and physical activity, or constrain unhealthy habits, such as drinking and smoking. Lindstrom et al. (2001) argue that social interaction may influence leisure-time physical activity through peer-pressure mechanisms. For example, jogging with a friend or joining a football team may make physical exercise less boring and painful, thus providing incentives to increase fitness and to keep weight under control. According to Haughton McNeill et al. "through social networks individuals form a sense of attachment and connectedness to one another providing access to resources and material goods that support physical activity (e.g., provision of child care services)" (2006, 1014). Folland (2007) argues that sympathetic relationships can provide "coaching" practices similar to those practised by fitness trainers. These hypotheses have been tested in several empirical studies. One of the main findings of this literature is the existence of a significant and positive correlation between social participation and physical activity (Brennan et al. 2003; Giles-Corti and Donovan 2003; Huston et al. 2003). On the other hand, individuals with poor social interactions generally show a higher propensity for unhealthy behaviors (Vander Wal, 2012).

4) *"Buffering effect"*. Social interactions and community cohesion provide moral and affective support which mitigates the psychological distress related to sickness. This "buffering effect" may play a role in improving patients' ability to recover, thereby improving the health status of sick people. The buffering effect of a cohesive network or community also works for healthy people by preventing depression and mental disorders often related to social isolation and acting as a source of self-esteem and mutual respect (Kawachi et al. 1999). Evidence of the buffering effect is also provided by the growing body of studies on the relationship between volunteering and health (Post 2005; Scheffler et al. 2007; Borgonovi 2008; Barron et al. 2009; Haski-Leventhal 2009).

The four arguments exposed above received support from several empirical studies finding significant and positive relationships between indicators of social capital and self-reported health. The two papers that are most closely related to ours in this literature employed instrumental variables estimates to determine the causal effect of social capital on health. D'Hombres et al. (2010) used data drawn from the 2001 Living Conditions, Lifestyles and Health survey collected in a sample of eight former Soviet Republics to analyse the effect of individual-level structural and cognitive social capital on health. The authors instrument their measures of social capital – i.e. social trust, membership in associations, and self-reported loneliness – with community-based instruments given by the heterogeneity in religious beliefs, in education, and in economic well-being, and the community-level social capital, as measured by the average levels of social trust, membership and loneliness. The study finds that trust is significantly and positively correlated with health, while loneliness has a significantly negative effect. The role of membership, on the other hand, remains unclear. In the other study, Kim et al. (2011) used survey data from the European and World Values Survey to estimate the contextual effects of country-level social trust on individual self-rated health. The authors instrument country-level social trust with corruption, religious fractionalisation, and population density measured at the country-level, to find that that increases in average country-level social trust lead to improved self-rated health.

In our empirical analysis we follow D'Hombres et al. (2010) and Kim et al. (2011) and use an instrumental variables approach to identify the effect of individual structural social capital on self-reported health.

*Theoretical background*

To provide a theoretical background for the relationship between social relations and health we refer to the model of health production developed by Contoyannis and Jones (2004). We assume that an individual's health is produced as follows:

$$H = h(C, SC, X, e) \qquad (1)$$

Where *H* is a measure of individual health, *C* is the set of consumption; *SC* represents individual structural social capital, *X* and *e* are the set of observable and unobservable personal characteristics, respectively.

In light of the arguments outlined above, we expect to find a significant and positive relationship between self-perceived health (SPH) and individual structural social capital.

$$SPH = f(C, \underset{+}{SC}, X, e) \qquad (1.1)$$

**3. Methodology and data**

Our empirical strategy is in two steps. First, the empirical model of perceived health can be represented through the following model:

$$H_i^* = \gamma SC_i + X_i'\beta + \varepsilon_i \qquad (2)$$

Where *H* is self-reported health for individual *i*; *SC* is our indicator of individual structural social capital; the *X* vector consists of annual household income and the other variables that are believed to influence self-perceived health; γ and β are the parameter to be estimated and ε is a random-error term.

We do not observe the "latent" variable $H_i^*$ in the data. Rather, we observe $H_i$ as a binary choice that takes value 1 (good or very good perceived health) if $H_{it}^*$ is positive and value 0 otherwise. Thus, the structure of (2) makes it suitable for estimation as a probit model:

$$\Pr(H_i = 1) = \Phi(\gamma SC_i + X_i'\beta_i) \qquad (3)$$

where Φ(·) is the cumulative distribution function of a normal standard.

A simple alternative to probit model is the linear probability model:

$$\Pr(H_i = 1) = \gamma SC_i + X_i'\beta_i \qquad (4)$$

In the second step, individual structural social capital is endogenous. Hence, we use the following latent model:

$$H_i^* = \gamma SC_i + X_i'\beta + \varepsilon_i \tag{5}$$

$$SC_i^* = Z_i'\pi + X_i'\beta + \mu_i \tag{6}$$

$H_i = 1$ if $H_i^* > 0$, 0 otherwise

$SC_i = 1$ if $SC_i^* > 0$, 0 otherwise  (7)

Where $Z_i'$ is the vector of additional instrumental variables that affect $SC_i$ but can be excluded from (5) as they do not directly affect $H_i$.

We estimate (5) – (7) using

- Instrumental variable (IV) probit estimator in which we assume that ($\varepsilon_i$, $\mu_i$) are jointly normally distributed, i.e. $(\varepsilon_i, \mu_i) \approx N(0, \sigma_{ij})$ and correlation ρ.

- Two-stage least-square (2SLS) estimator in which the binary nature of dependent variable is ignored and heteroskedasticity robust standard errors are used.

- Bivariate probit model (Maddala 1983, 122-123) in which

$$\Pr(H_i = 1, SC_i = 1) = \Phi_2(\gamma SC_i + X_i'\beta_i, Z_i'\pi + X_i'\beta_i, \rho) \tag{8}$$

Where $\Phi_2(\cdot)$ is the cumulative distribution function of a bivariate normal standard with zero means, unit variances and correlation ρ.

Individual structural social capital is measured through the habit of meeting friends. This indicator is drawn from the 2000 wave of the Multipurpose Survey on Household (MSH) conducted by the Italian Institute of Statistics (Istat). This survey investigates a wide range of social behaviours and perceptions by means of face-to-face interviews on a nationally and regionally representative sample of 24,000 households, roughly corresponding to 50,000 individuals. Since the MHS does not collect information on household income, we merged this source with the Survey on Household Income and Wealth (SHIW) carried out by the Bank of Italy through a statistical matching procedure. Basically, the household income of an individual from the SHIW has been imputed to a similar individual from the MHS through a regression imputation with random residuals (see Appendix 1 for further details).

After deleting observations with missing data on the dependent variable and on the indicator of social capital, the final dataset is a cross-section of 46,868 observations containing information on individual behaviours and perceptions as well as on household income.

Our dependent variable is self-perceived good health, as measured by a dummy that is equal to 1 if the respondent reports good or very good health. The main independent variable in the analysis is individual structural social capital. This is measured through a binary indicator of the frequency of meetings with friends, which is coded as 1 if the interviewee meets friends everyday or more times a week.

In order to account for other phenomena which might influence health and social capital, we include in the analysis a set of individual and household control variables.

At the individual level, we account for gender (female), marital status (single, divorced and widowed), age (dummies 31-40, 41-50, 51-65, older than 65), education (elementary, junior high school, high school education and undergraduate degree or more), reading newspapers and work status (unemployed, self-employed, retired, student). Moreover, we measure the quality of the surrounding social environment through an indicator of the subjective perception of its safety. At the household level, we control for the natural logarithm of the imputed household income (sum of labour income, capital income and pensions) obtained through the statistical matching procedure. In addition, we account for family size, number of children in three age groups (dummies 0-5, 6-12, 13-17), homeownership and the characteristics of homes (whether it is council house or not). Finally, we also control for the size of municipality. All the variables are described in detail in Table B1 in Appendix 2. Summary weighted statistics are reported in Table 1. On average, 71% of respondents report good or very good health. 70% meet friends at least twice a week. Over half of respondents are female and married. Over half of the sample report low education (elementary and junior high school) while only 8 % hold an undergraduate degree. The largest group of individuals (22 %) is aged between 51 and 65, followed (21%) by individuals aged 18 to 31 (reference group).

Over half of respondents have children aged between 0 and 17. 72 % of respondents are homeowners, while 62 % live in a popular house.

| Table 1. Descriptive statistics | | |
|---|---|---|
| | Mean | St. dev. |
| *Dependent variable* | | |
| Self-perceived good health | 0.71 | 0.45 |
| *Structural social capital* | | |
| Meetings with friends | 0.70 | 0.46 |
| *Instrumental variables* | | |
| Mass attendance | 0.25 | 0.43 |
| Meetings with friends at the community level (average) | 0.71 | 0.03 |
| *Demographic and socio-economic characteristics* | | |
| Female | 0.52 | 0.50 |
| Single | 0.26 | 0.44 |
| Divorced | 0.04 | 0.20 |
| Widowed | 0.09 | 0.29 |
| Age31-40 | 0.19 | 0.39 |
| Age41-50 | 0.17 | 0.37 |
| Age51-65 | 0.22 | 0.42 |
| Age > 65 | 0.20 | 0.40 |
| Household size | 3.10 | 1.30 |
| Children 0-5 | 0.13 | 0.40 |
| Children 6-12 | 0.17 | 0.46 |
| Children 13-17 | 0.15 | 0.41 |
| Elementary | 0.25 | 0.43 |
| Junior high school | 0.28 | 0.45 |
| High school (diploma) | 0.32 | 0.47 |
| Undergraduate degree and beyond | 0.08 | 0.27 |
| Household income (ln) | 10.77 | 0.44 |
| Self-employed | 0.11 | 0.31 |
| Unemployed | 0.05 | 0.23 |
| Retired | 0.23 | 0.42 |
| Student | 0.05 | 0.23 |
| Newspaper reader | 0.24 | 0.42 |
| Homeowner | 0.72 | 0.45 |
| Civil house | 0.62 | 0.49 |
| Micro-criminality | 0.03 | 0.17 |
| *Size of municipality* | | |
| Metropolis | 0.17 | 0.37 |
| Neighbouring metropolis | 0.08 | 0.27 |
| 2000-10000 | 0.27 | 0.44 |
| 10000-50000 | 0.23 | 0.42 |
| >50000 | 0.16 | 0.37 |

*3.1 Instrumental variables*

The reliability of probit estimates may suffer from the endogeneity problems described in the introduction, which suggest caution in interpreting correlations as causal relationships. We try to circumvent endogeneity problems by instrumenting the frequency of meetings with friends. As pointed out by French and Popovici (2011), a reliable instrumental variable must meet at least two criteria. First, it must be theoretically justified and statistically correlated with individual structural social capital ("relevance" condition), after controlling for all other exogenous regressors. Second, it must be uncorrelated with the disturbance term of the health equation ("orthogonality" condition). We selected from our data source the following two theoretically convenient and econometrically valid instrumental variables (IVs):

- Mass attendance, as given by a binary variable coded as 1 if the interviewee attends a Mass once a week.

- Meetings with friends at the community level, given by the average frequency with which people meet friends at the community level. The reference group of individuals is the group of people at the municipality level in the same age group and at the same education level. We consider six categories of municipality size (metropolis, neighboring metropolis, less than 2000 inhabitants, 2,000-10,000, 10,000-50,000, more than 50,000 inhabitants), two age groups (less than 40 years, more than 40 years) and two education levels (low, high). Thus we have 24 reference groups in each of the 20 Italian regions. The wealth of ties in the local community is calculated as the mean value of the daily frequency of meetings with friends for each of the 24 reference groups in each of the 20 Italian regions. We obtain 480 combinations across which the 46,868 observations of the sample are distributed[6].

---

[6] The MHS sample is representative at the national and regional level, as well as at the level of 6 possible categories of municipality. Categories include: A) municipalities belonging to a metropolitan area, separated into: 1) municipalities in the centre of a metropolitan area. These are Bari, Bologna, Cagliari, Catania, Firenze, Genova, Milano, Napoli, Roma, Torino, Venezia. 2) Municipalities immediately around those metropolitan areas. B) Municipalities outside of metropolitan areas, separated into: 3) Municipalities with a population of under 2,000. 4) Municipalities with a population of between 2,001 and 10,000. 5) Municipalities with a population of between 10,001 and 50,000. 6) Municipalities with more than 50,000 inhabitants.

The relevance condition is directly testable by regressing individual structural social capital on the IVs and all other exogenous variables from the structural equation. The first stage of our IV regressions shows that both the instruments are strongly correlated with the endogenous variable (see Section 4 and Appendix 4 Table D1).

The positive relationship between community-level social capital and the individual consumption of relational goods[7] has been already documented in the theoretical and empirical literature (Gui and Sugden 2005; Antoci et al. 2007). If the social environment is rich in participation opportunities, because many people already participate and there are well-established networks of relations, then the time individuals spend on social interactions will be more rewarding: as a result, people will be stimulated to meet friends more frequently (Antoci et al. 2012). By contrast, if the surrounding environment is relationally poor (i.e. the average level of participation is low), individuals may be forced to replace human interactions with private consumption (e.g. playing a virtual match against the computer instead of meeting friends on a sport field, or chatting with unknown and distant people through the web instead of talking with neighbours). As a result, people may be discouraged from meeting others and are more likely to report being socially isolated.

As for Mass attendance, several studies suggested that forms of religious participation may promote and sustain regular opportunities for social interaction among persons with common values and interests (Ellison and George 1994; Ellison 1995; Soydemir et al. 2004). Religious participation usually takes place in a group context and this involves social relationships and the formation of networks, i.e. social capital (Hawe and Shiell 2000; George et al. 2002; Chiswich and Mirtcheva 2013).

---

[7] Relational goods are a distinctive type of good that can only be enjoyed if shared with others. They are different from private goods, which are enjoyed alone (Uhlaner 1989). A peculiarity of relational goods is that it is virtually impossible to separate their production from consumption, since it is very likely that they will coincide (Gui and Sugden 1995). For example, a football match with friends is enjoyed (consumed) in the very moment of its production (i.e. the 90 minutes spent on the sports field). For the sake of simplicity, the frequency of meetings with friends, the frequency of relational goods consumption, and individual-level structural social capital can be considered equivalent/synonymous in this discussion.

The orthogonality condition cannot be tested directly as it involves a relationship between the instruments and the error term of the structural equation. Hence, we rely on the following theoretical considerations and intuitions and we indirectly address the validity condition through an andogeneity test, an over-identification test and a weak instrument test we present in Section 4.

As regards the community-level instrument, it must be stated that several studies report a positive correlation between community social capital and individual health (Kawachi et al. 1999; Islam et al. 2006; 2008). However, as properly reported by D'Hombres et al. (2010), these studies do not simultaneously include indicators of individual social capital in the health equation. "Thus, the effect of community level social capital can be due to its positive correlation with individual social capital" (D'Hombres et al. 2010, 62). Many authors show that the effect of community-level social capital becomes insignificant after controlling for measures of individual-level social capital (Poortinga 2006a, 2006b; Subramanian et al. 2002), or other individual-level socio-demographic characteristics (Kennelly et al. 2003). D'Hombres et al. (2010) conclude that community social capital does not have any independent effect on self-reported health once individual social capital is controlled for and use community-level social capital to instrument individual social capital to explain the determinants of self-reported health in eight countries from the Commonwealth of Independent States. In following the same empirical strategy, we provide further support to the validity of the instrument picked by the authors in a representative sample of the Italian population.

It is worth remembering that the community-level variable we account for in the analysis captures the frequency with which friends spend time together (e.g. for going out for dinner or playing a football match). This measure does not represent the community's civic engagement, which may, instead, have a role in promoting public health. Rather, in Italian society, this specific form of social capital has been found neither to have a significant correlation with membership in associations (Degli Antoni 2009; Felice 2013), nor with economic and political outcomes such as the quality of public services and public health infrastructures (Sabatini 2008; 2009).

As for Mass attendance, we are aware that a number of studies put forward a positive association between physical health, mental health, mortality and religious participation (Ellison 1995; Koening et al. 1999). According to Soydemir et al. (2004) attending religious services either reduces the number of illnesses or increases positive health outcomes. The authors use three arguments to interpret their findings. First, Mass attendance may provide regular opportunities for social interaction among people with common values and interests. Second, religious groups may exercise social control over their members' lifestyles through specific norms that promote healthy behaviours across various life domains (Ellison and George 1994). Third, Mass attendance involves patterned engagements in ritual events to which participants assign a special significance. Meanings attached to these ritualistic events may promote feelings of well-being fostering positive mental health as well as identifying with religious scriptures and texts (Wikstrom 1987; Umberson 1987). However, these studies are not able to empirically disentangle the role of these possible channels of transmission. What has been found so far in the empirical literature is a correlation between some forms of religious participation and specific health outcomes. We argue that this correlation is heavily dependent on the specific characteristics of the religious groups involved and on the historical, social, and relational circumstances in which participation takes place. In Italy, Mass attendance is generally not related to the involvement in forms of active religious participation. Rather, it is a weekly ritual that seems hardly capable to influence attendants' everyday lifestyle. Differently from what may happen in the course of active religious participation, during Catholic Masses individuals do not receive detailed instructions on how to behave in their daily life. Rather, these religious services just provide the devoted with simple moral guidelines. An in-depth account of this way of religious participation is given in the two main historical studies on the Italian society from a sociological perspective, carried out by Banfield (1958) and Putnam et al. (1993). In describing the role of the Catholic Church in the Italian society, Banfield (1958: 129-130) wrote: "How men do behave and how they should behave are different matters … they get little religion instruction. A peasant grandmother tells her grandchildren the stories of miracles and sacred things she heard from her grandmother. At six a child learns his catechism, a meagre list of questions and answers which is likely to be forgotten soon. In later life the individual who goes to church hears simple sermons. The priest says, for example, that to be a good catholic one must love God, obey the laws of the church, and do right. This is the extent of the ordinary person's religious training."

In *Making Democracy Work*, Putnam and colleagues (1993) document that all manifestations of religiosity and clericalism - such as attendance at Mass and religious (as opposed to civic) marriages – are negatively correlated with civic engagement[8]. The authors describe the Catholic Church as a hierarchical institution whose leadership sees citizen engagement as a potential threat to its privileged role in Italian political and social life. In the context of such a rigidly hierarchical relationship between clerical institutions and their devoted, it is difficult to imagine that going to Masses can actually influence the attendants' daily behavior in an healthy or unhealthy way. Overall, the attendance of Catholic Masses might thus neither be related to forms of social control discouraging unhealthy behaviors, nor to substantial increases in subjective well-being in turn alleviating mental distress. The empirical evidence on Italy systematically supports this argument. Using Italian survey data, Fiorillo and Sabatini (2011) and Fiorillo and Nappo (2014) found no evidence of a significant association between Masses attendance and self-perceived health. To further test this argument, we regressed a number of indicators of unhealthy behaviors on Masses attendance and on the set of covariates also included in our health equation. We found no significant relationship between religious participation and the habits of drinking excessive amounts (more than one glass per day) of various types of alcoholic drinks (e.g. wine, beer, or aperitifs), ,. There also is no relationship between Mass attendance and indicators of a healthy diet, such as the habit of drinking one liter of water (or more) a day. Estimates are reported in Appendix 3 Table C1.

---

[8] Note that this result is not inconsistent with our finding that Mass attendance is significantly and positively correlated with the frequency of meetings with friends, as Putnam et al. (1993) actually refer to "civic engagement" as forms of associational activity and active participation for the common good. In Putnam's work it is generally acknowledged that the Catholic Church contain numerous opportunities for horizontal, informal, engagement for recreational purposes.

The inability to exert effective social control could has also been explained as a result of the loss of credibility related to the tendency of clerical institutions to bury cases of child sex abuse (citations) and to officially endorse political candidates with considerably profligate lifestyles such as, for example, the former Prime Minister Silvio Berlusconi (Sabatini 2012), despite he was found guilty of paying for sex with an underage prostitute in the course of a dissolute nightlife (which looks far from being a healthy behavior).

The inconsistency between these findings and the literature reporting a positive correlation between religion and health may be due to the fact that, in previous studies, health outcomes were studied in other countries, in relation to other rituals (e.g. protestant cults), or without distinguishing between rituals. O'Reilly and Rosato (2008) used census-based longitudinal data with five-years follow-up to conduct one of the rare comparative studies examining variation in overall and cause-specific mortality by religious affiliation in Northern Ireland. The authors found robust evidence that Catholics had higher mortality than non-Catholics. The significance of this difference, however, disappeared after adjusting for socio-economic status. In the fully adjusted models, it was the affiliation with protestant groups that was significantly and positively correlated with better health outcomes, due to the lower mortality from alcohol-related deaths and lung cancer. Roman Catholic Church's limited ability (or will) to exert social control is also suggested by studies on the relationship between religious beliefs and portfolio choices. Using U.S. data on church membership and portfolio choices for the period 1980-2005, the authors found that gambling propensity is significantly stronger in areas with higher concentrations of Catholics relative to Protestants.

The findings briefly mentioned above seem to be overall consistent with Putnam's view that the Roman Catholic Church is more interested in political control, than in controlling the lifestyle of its devoted. In another comparative study, Klanjšek et al. (2012) found that not every type of religiosity has a buffering effect on deviance: if one's religiousness is predominately extrinsic, then its inhibiting effect is weak or does not exist. According to Allport and Ross (1967), a religious orientation is extrinsic when observant tend to use religion for their own ends, e.g., as a source of solace and security, as a mean to socialize, to affirm one's social status, to develop social connections to be exploited for the sake of their particular interests. Historical studies in sociology suggest that religious participation, in fact, also has relevant extrinsic motivations in Italy. According to Banfield's (1958: p. 131) analysis of Italian backward Southern areas, "The relation between the believer and God is characteristically based on the *interesse* (interest) of each. One party wants to be honored with candles and masses. The other wants protection" and sees the Catholic Church as a temporal institution that may be exploited to reach extrinsic goals, e.g. to create connections with people in position of power and gain favors, or for leisure and recreation (e.g. as a place where to meet friends). This argument about the extrinsic motives of Mass attendance is consistent with our finding that the latter is significantly and positively correlated with self-reported economic well-being (Table C2). We also regressed respondents' satisfaction with their relationships with friends and with their leisure time on our instrument and all the other covariates employed in the analysis. Our estimates that these variables are significantly and positively correlated with Mass attendance are consistent with the argument that religious participation is also seen as a means to preserve and develop social contacts. Thus, it seems reasonable to argue that any significant influence of Mass attendance on self-reported health might be mediated by meetings with friends, i.e. the endogenous variable in our health equation.

Taken together the arguments exposed above suggest that there is no direct link to self-reported health from community-level frequency of meetings with friends and Mass attendance per se in our sample.

## 4. Empirical results and discussions

*Determinants of self-perceived good health*

Table 2 presents estimates of the health equation (2). To compare relative magnitudes of the effects of the independent variables, we report their marginal effects. Column 1 reports the probit estimate (equation 3) and column 2 presents the linear estimation (equation 4).

Before discussing the impact of individual structural social capital, we briefly present the effect of individual and household variables on self-perceived good health. As the estimates resulting from probit and linear specifications are almost identical, we base the discussion below only on the results displayed in column 1 of Table 2.

Regarding individual characteristics, we find that women and men show statistically significant difference in good health, with women turn out lower levels of perceived good health. Education is a relevant predictor of health. Having a high-school leaving certificate increases the probability of perceived good health. This probability further rises in individuals with an undergraduate or graduate degree. Results on gender and education effects are in line with the findings of Portinga (2006a), Iversen (2008), Cooper et al. (2006) and Etile (2014). As expected and found in other countries (Petrou and Kupek 2008; D'Hombres et al. 2010; Ronconi et al. 2010), age is negatively correlated with good health.

The household characteristics are important predictors of health. Being single and divorced decrease the probability of reporting good health, while household size rises the probability of declaring good health. People with children aged between 0 and 5 present a higher likelihood of reporting good health.

Table 2. Probit and least-squares estimates

|  | I | | II | |
|---|---|---|---|---|
|  | Probit | | Least squares | |
|  | Marg. Eff. | Std. Err. | Marg. Eff. | Std. Err. |
| Meetings with friends | 0.048*** | 0.005 | 0.044*** | 0.005 |
| Female | -0.013*** | 0.005 | -0.014*** | 0.004 |
| Single | -0.122*** | 0.008 | -0.110*** | 0.007 |
| Divorced | -0.031*** | 0.012 | -0.025** | 0.011 |
| Widowed | 0.007 | 0.008 | 0.006 | 0.009 |
| Age31-40 | -0.049*** | 0.009 | -0.042*** | 0.007 |
| Age41-50 | -0.091*** | 0.011 | -0.077*** | 0.008 |
| Age51-65 | -0.161*** | 0.012 | -0.138*** | 0.009 |
| Age > 65 | -0.356*** | 0.014 | -0.335*** | 0.012 |
| Household size | 0.021*** | 0.003 | 0.024*** | 0.002 |
| Children 0-5 | 0.036*** | 0.008 | 0.013*** | 0.005 |
| Children 6-12 | - 0.000 | 0.006 | -0.007* | 0.004 |
| Children 13-17 | 0.036*** | 0.006 | 0.022*** | 0.005 |
| Elementary | 0.053*** | 0.008 | 0.068*** | 0.010 |
| Junior high school | 0.093*** | 0.010 | 0.118*** | 0.011 |
| High school (diploma) | 0.110*** | 0.011 | 0.132*** | 0.012 |
| Bachelor's degree and beyond | 0.130*** | 0.011 | 0.159*** | 0.014 |
| Household income (ln) | 0.071*** | 0.010 | 0.065*** | 0.009 |
| Self-employed | 0.021*** | 0.007 | 0.018*** | 0.006 |
| Unemployed | -0.030*** | 0.011 | -0.025*** | 0.009 |
| Retired | -0.040*** | 0.007 | -0.043*** | 0.007 |
| Student | 0.023** | 0.011 | 0.019** | 0.009 |
| Newspaper reader | 0.024*** | 0.005 | 0.022*** | 0.005 |
| Homeowner | -0.030*** | 0.006 | -0.029*** | 0.005 |
| Civil house | 0.005 | 0.004 | 0.005 | 0.004 |
| Micro-criminality | -0.003 | 0.013 | -0.003 | 0.012 |
| *Size of municipality* | | | | |
| Metropolis | -0.015* | 0.009 | -0.013 | 0.008 |
| Neighbouring metropolis | -0.018 | 0.011 | -0.015 | 0.009 |
| 2000-10000 | -0.006 | 0.008 | -0.005 | 0.007 |
| 10000-50000 | -0.017* | 0.009 | -0.014* | 0.008 |
| >50000 | -0.010 | 0.009 | -0.008 | 0.008 |
| Regional dummies | Yes | | Yes | |
| Pseudo R-squared | 0.13 | | 0.16 | |
| Log-likelihood | -23148.97 | | | |
| No. of observations | 44954 | | 44954 | |

Note: The dependent variable *Self-perceived good health* is a binary variable (1 = good and very good, 0 otherwise).. See Appendix 2 Table 1B for a detailed description of regressors. Regional dummies are omitted for space reasons. Standard errors are corrected for heteroskedasticity. The symbols ***, **, * denote that the coefficient is statistically different from zero at 1, 5 and 10 percent.

This finding seems to support the hypotheses on the "relational" incentives towards healthy behaviour: as noted by Folland, "responsibility to others requires at a minimum that one stay alive and healthy" (2007, 2345) and can discourage potentially self-damaging behaviours such as excessive drinking and smoking. As expected, the (imputed) household income is significantly and positively correlated with good health (as in Borgonovi, 2008; Theodossiou and Zangelidis 2009; Yamamura 2011).

Work status is found to be another important explanatory variable. In the literature, unemployment was identified as a key socioeconomic determinant of health (see Cooper et al. 2006). We find that being unemployed increases the individuals' probability of rating their own health as poor by about 3%. By contrast, self-employed workers exhibit a 2.1 points higher probability of reporting good health. Research into the relationship between unemployment and well-being generally agree that people in secure employment recover more quickly from illness (Bartley 1994; Dorling 2009). In contrast, unemployment increases the chance of being ill, especially for those who had never worked or had had poorly paid jobs (Gerdtham and Johannesson 2003; Bartley et al. 2004; Koziel et al., 2010). Unemployment increases rates of depression, particularly in the young (Branthwaite and Garcia 1985; Artazcoz et al. 2004) and causes unhappiness (Clarke and Oswald 1994), which has, in turn, been linked to poor health (Danner et al. 2001; Bjørnskov 2008; Veenhoven 2008). Finally, interesting results regard the status of student and retired, which are significantly and negatively correlated with perceived good health, respectively, at 2.3% and 4%, and the habit of reading newspapers every day, which is significantly and positively correlated with good health. Previous cross-sectional studies (e.g., Bosse et al. 1987; Butterworth et al. 2006) have found a positive association between retirement and the worsening of mental health, which may in turn influence the self-reported health conditions of individuals. Drawing on the 1996 wave of the European Community Household Panel, van Doorslaer and Koolman (2004) find that retired workers are more likely to exhibit worse self-reported health. However, other authors (e.g., Herzog et al. 1991; Ross and Drentea, 1998) did not find any significant correlation, and these discrepancies have not been resolved by longitudinal investigations (Butterworth et al. 2006). Further research is certainly needed on this topic to better understand how the effect of retirement and of the status of being a student on psychological and physical health varies with age and if it may be influenced by the social norms surrounding employment status at different ages.

Overall, results from our estimates show the existence of health disparities based on socio-economic status in Italy, as already claimed by two previous studies (Atella et al. 2004; Masseria and Giannoni 2010). Even though the Italian NHS is in principle designed to provide universal coverage for all citizens at the point of use, poorer and less educated individuals are more likely to report poor health conditions. The risk is even worse for unemployed and retired workers. The significance of regional dummies (not reported) also reveals the existence of relevant territorial health disparities. This result may reflect the influence of a number of local factors and suggests the need for a regional analysis of the socio-economic determinants of health, which should draw attention to the role of public policies. The Italian healthcare system is in fact going through a major transition, affecting policy decisions, financing methods and service provision. These changes are taking place within the larger context of the so-called "devolution", a process of decentralization, which has afforded regions greater autonomy in the definition of health policies, including the responsibility of financing healthcare through regional taxes and of allowing for-profit providers to replace the NHS in the provision of a growing number of healthcare services. Some authors have underlined how this decentralization process implies a substantial risk of exacerbating the incidence of health inequalities (De Vries 2000; Walker 2002; Mosca 2006).

Since our main interest is the impact of individual structural social capital on self-perceived health, we now turn to the analysis of the marginal effects of the measure of meetings with friends, focusing on the estimates displayed in Table 2. In line with our hypothesis, individual structural social capital is found to be strongly and positively associated with self-perceived good health, irrespective of the estimation procedure. Individuals who meet friends every day or more times a week are 4.8% more likely to report self-perceived good health. Meetings with friends are likely to facilitate the transfer of health-related information and to reduce psychological distress. Moreover, networks of friends may provide mutual assistance mechanisms and foster healthy behaviours (as indicated in Section 2).

*Self-perceived good health and individual structural social capital: IV estimates*

However, because of the statistical problems we discussed in the previous Section, we must be careful in interpreting this correlation as causal. In order to shed more light on the causal relationship connecting individual structural social capital to perceived health, we now turn to instrumental variables estimates. In this Section, we will rely on two-stage least-squares estimates since the results presented above show that probit and linear specifications give very similar results. We use bivariate probit estimates as a robustness check.

Results are reported in Table 3, Panel A, B and C. Panel A, presents the marginal effect of individual structural social capital as well as of all covariates on self-perceived good health. Panel B and C of the Table 3 reports diagnostic tests of the validity of our instrumental variable estimators. Columns 1 and 3 report results of two-stage least-squares estimates (IV Probit and 2SLS). As a robustness check, in column 2 we report the results of the bivariate probit model. Panel A shows that in two steps estimates (IV probit and 2SLS) the sign and significance of all covariates is confirmed (also see Table 2). In addition, IV probit's results are consistent with those of the bivariate probit. The negative sign of $\rho$, $(\hat{\rho} = -0.141)$ indicates a negative correlation between $\varepsilon_i$ and $\mu_i$. This suggests that the unobservable factors that raise the likelihood to meet friends also lower the likelihood of reporting good health, conditional to all the other regressors included in the health equations. In Panel B, we report the test of exclusion restrictions of the instrumental variables from the health equation. The null hypothesis of non-significance of the two instruments cannot be rejected at the conventional 5 per cent level. Finally, the first condition is directly tested by regressing the endogenous variable, "meetings with friends", on the instruments – Mass attendance and meetings with friends at the community level – and on all the other exogenous

| Table 3. Panel A. Instrumental variables estimates | | | | | | |
|---|---|---|---|---|---|---|
| | I | | II | | III | |
| | IV Probit | | Biprobit | | 2SLS | |
| | Marg. Eff. | Std. Err. | Marg. Eff. | Std. Err. | Marg. Eff. | Std. Err. |
| Meetings with friends | 0.162*** | 0.054 | 0.110*** | 0.038 | 0.126*** | 0.044 |
| Female | -0.006 | 0.005 | -0.063*** | 0.005 | -0.009* | 0.005 |
| Single | -0.133*** | 0.009 | -0.024*** | 0.008 | -0.119*** | 0.008 |
| Divorced | -0.030** | 0.012 | -0.023* | 0.012 | -0.024** | 0.011 |
| Widowed | 0.008 | 0.008 | 0.001 | 0.009 | 0.007 | 0.009 |
| Age31-40 | -0.040*** | 0.010 | -0.125*** | 0.010 | -0.036*** | 0.007 |
| Age41-50 | -0.074*** | 0.013 | -0.062*** | 0.013 | -0.066*** | 0.010 |
| Age51-65 | -0.138*** | 0.016 | -0.143*** | 0.013 | -0.122*** | 0.012 |
| Age > 65 | -0.319*** | 0.023 | -0.314*** | 0.014 | -0.312*** | 0.017 |
| Household size | 0.022*** | 0.003 | 0.014*** | 0.003 | 0.025*** | 0.002 |
| Children 0-5 | 0.043*** | 0.008 | -0.022*** | 0.007 | 0.018*** | 0.006 |
| Children 6-12 | 0.000 | 0.006 | -0.011* | 0.005 | -0.007 | 0.004 |
| Children 13-17 | 0.036*** | 0.006 | 0.030*** | 0.007 | 0.022*** | 0.005 |
| Elementary | 0.048*** | 0.009 | 0.079*** | 0.009 | 0.065*** | 0.010 |
| Junior high school | 0.086*** | 0.010 | 0.123*** | 0.011 | 0.113*** | 0.012 |
| High school (diploma) | 0.103*** | 0.012 | 0.124*** | 0.012 | 0.128*** | 0.013 |
| Bachelor's degree and beyond | 0.123*** | 0.012 | 0.145*** | 0.014 | 0.153*** | 0.015 |
| Household income (ln) | 0.074*** | 0.010 | 0.024** | 0.010 | 0.068*** | 0.009 |
| Self-employed | 0.020*** | 0.007 | 0.027*** | 0.007 | 0.017*** | 0.006 |
| Unemployed | -0.031*** | 0.011 | -0.004 | 0.012 | -0.025*** | 0.009 |
| Retired | -0.048*** | 0.008 | 0.002 | 0.007 | -0.049*** | 0.008 |
| Student | 0.018*** | 0.012 | 0.111*** | 0.014 | 0.015 | 0.009 |
| Newspaper reader | 0.018*** | 0.006 | 0.056*** | 0.006 | 0.017*** | 0.005 |
| Homeowner | -0.034*** | 0.006 | 0.007 | 0.006 | -0.032*** | 0.006 |
| Civil house | 0.005 | 0.004 | 0.005 | 0.005 | 0.004 | 0.004 |
| Micro-criminality | -0.007 | 0.013 | 0.005 | 0.014 | -0.006 | 0.012 |
| *Size of municipality* | | | | | | |
| Metropolis | -0.017* | 0.009 | -0.014 | 0.010 | -0.014* | 0.008 |
| Neighbouring metropolis | -0.017 | 0.011 | -0.017 | 0.011 | -0.015 | 0.009 |
| 2000-10000 | -0.005 | 0.009 | -0.010 | 0.009 | -0.004 | 0.008 |
| 10000-50000 | -0.017* | 0.009 | -0.017 | 0.010 | -0.014* | 0.008 |
| >50000 | -0.010 | 0.009 | -0.010 | 0.010 | -0.008 | 0.008 |
| Regional dummies | Yes | | Yes | | Yes | |
| ρ | -0.141 | 0.065 | -0.164 | 0.080 | | |
| Pseudo R-squared | | | | | 0.16 | |
| Log-likelihood | -48427.42 | | -47057.11 | | | |
| No. of observations | 44704 | | 44704 | | 44704 | |

Note: The dependent variable *Self-perceived good health* is a binary variable (1 = good and very good, 0 otherwise). Instruments for meetings with friends are meetings with friends at the community level and *Mass attendance*. Standard errors are corrected for heteroskedasticity. The symbols ***, **, * denote that the coefficient is statistically different from zero at 1, 5 and 10 percent.

| Table 3. Panel B. Instrumental variable estimates: tests | I | II | III |
|---|---|---|---|
| | IV Probit | Biprobit | 2SLS |
| Test of exclusion of instruments from Health equation | | | |
| *F*-statistics | | | 2.01 |
| *Chi2*-statistics | 5.06 | | |
| *P*-value | 0.07 | | 0.13 |
| First stage: joint significance of instruments | | | |
| *F*-statistics | | | 233.70 |
| *Chi2*-statistics | 412.21 | | |
| *P*-value | 0.00 | | 0.000 |
| Test of endogeneity | | | |
| Robust Durbin-Wu-Hausman test | | | |
| *Chi2*-statistics | | | 3.43 |
| Wald test | | | |
| *Chi2*-statistics | 4.57 | 4.07 | |
| *P*-value | 0.03 | 0.04 | 0.06 |
| Overidentification test | | | |
| Sargan test | | | |
| *Chi2*-statistics | | | 0.64 |
| Amemiya-Lee-Newey | | | |
| *Chi2*-statistics | 0.51 | | |
| *P*-value | 0.48 | | 0.42 |

Table 3. Panel C. Stock and Yogo weak identification tests

| Minimum eigenvalue statistic: 230.59 Critical values Ho: Instruments are weak | | | | |
|---|---|---|---|---|
| | 5% | 10% | 20% | 30% |
| 2SLS relative bias | (not available) | | | |
| | 10% | 15% | 20% | 25% |
| 2SLS Size of nominal 5% Wald test | 19.93 | 11.59 | 8.75 | 7.25 |

variables. The coefficients of instrumental variables are significantly different from zero at the conventional level of 5 percent (p-values 0.00). The F statistic for joint significance of the instruments in the first stage is 233.70.

The second condition is indirectly addressed by means of an endogeneity test and an over-identification test. The Wald tests and the robust Durbin-Wu-Hausman test lead to rejection of the null hypothesis that "meetings with friends" is exogenous at the conventional level of 5 percent. Since we use two instrumental variables for meetings with friends, we can conduct an over-identification test. The Amemiya-Lee-Newey test and the Sargan test of over-identifying restrictions do not lead us to reject the null hypothesis that the excluded instruments are valid instruments, i.e., uncorrelated with the error term, and that they are correctly excluded from the estimated equation, with a *p*-value, respectively, of 0.48 and 0.42.

To further test the strength of our instruments we also report the tests proposed by Stock and Yogo (2005). One commonly used diagnostic of weak instruments is the F statistic for joint significance of the instruments in the first stage of the endogenous variable on the instruments and all other exogenous variables. A widely used rule of thumb suggested by Staiger and Stock (1997) views an F statistic of less than 10 as indicating weak instruments. Using this rule, as the F statistic in the first stage is 233.70 well above the threshold of 10, we can conclude that our instruments are not weak. However, Stock and Yogo (2005) proposed two test approaches, under the assumption of homoskedastic errors that lead to critical values for F statistic. The first approach, applicable only if there are three or more overidentifying restrictions, suggests that the rule of thumb is reasonable. The second approach can lead to F statistic critical values that are much greater than 10 in models that are overidentified (see Stock and Yogo 2005, 95-103; Cameron and Trivedi 2010, 196-197).

Panel C of Table 3 reports the tests of Stock and Yogo. The first test is not available because the model is overidentified with only two instruments. Regarding the second test, if we are willing to tolerate a distortion for 5% Wald test based on the 2SLS estimator, so that the true size can be at most 10%, then we reject the null hypothesis if the test statistics exceeds 19.93. The reported minimum eigenvalue statistics of 230.59 equals the F statistics that the instruments are equal to zero if default standard errors are used in the first stage regression. We instead used robust standard errors, which led to F of 233.70. But both F statistics greatly exceed the critical values of 19.93, so we feel confident in rejecting the null hypothesis of weak instruments.

Taken together with the non-rejection of the tests of endogeneity, of over-identification, of not weak instruments and the theoretical considerations mentioned in the previous Section, this suggests that our set of instruments is reasonable.

The structural social capital variable is highly positively associated with self-perceived good health, and its marginal effect increases with respect to the result of Table 2 column 2: individuals who meet friends every day or more times a week are around 5% (Probit) to 16% (IV Probit) more likely to report self-perceived good health. Since the estimates now account for the endogeneity problems described in Section 3, we are more confident that this positive association can be interpreted as the result of a causal effect of structural social capital on self-perceived good health.

## 5. Conclusions

In this paper we have investigated the impact of individual structural social capital on individual self-perceived good health in Italy. Results of the empirical analysis support the hypothesis that individual structural social capital improves the health conditions of individuals. Following the previous literature, we argue that the mechanism may work through four main transmission channels, involving the diffusion of relevant health information, the establishment of mutual assistance mechanisms, the promotion of healthy behaviours, and the so-called buffering effect, i.e. the ability of community interactions to provide moral and affective support which mitigates the distress related to sickness.

In our view, this study makes a contribution to the literature along three substantive ways. First, it provides new relevant evidence of the role of social capital in health. To the best of our knowledge this is the first time the relationship between these two variables is assessed in a Mediterranean country. Second, it suggests a new, intermediate, objective for health policies: according to our results, policy makers should focus on social participation and social cohesion as powerful (and relatively inexpensive) means for the improvement of individual health. Third, our study attempts to move the literature forward away from merely descriptive analysis to causal analysis by trying a first step in the identification of the causal effect of individual-level social capital on health in Italy. The straightforward implication for future research is the need to better identify, both theoretically and empirically, to what extent and under which conditions each of the four suggested transmission mechanisms of the effect of social capital actually works. A further important implication is the need to investigate more in depth the sources of social capital and the causes of disparities in access to health information and healthcare services.

**Appendix 1. Statistical matching**

In simple terms, the matching procedure consists of the imputation of the household income of an individual from the SHIW to a similar individual from the Multipurpose Survey. As in Fiorillo (2008), let *A* be the MSH dataset (the so-called "base file") collecting information on $X_A$ variables for each of $n_A$ records, and let *B* be the SHIW dataset (the "supplemental file") comprising $X_B$ variables for each of $n_B$ records. Let $X = (X_1,...,X_P)$ be the vector of variables measured in both the files, i.e. for each of the units $n_A$ and $n_B$ included in the two datasets. The remaining variables in each of the files will be referred to as $Y = (Y_1,...,Y_Q)$ in file *A* and as $Z = (Z_1,...,Z_R)$ in file *B*. The statistical matching procedure is aimed at creating a file *C* which includes all the variables *X*, *Y*, and *Z* for each of $n_A$ records of the base file. For each unit in file *A* we identify a similar unit in file *B* as a function of the *X* "common" variables. After this, we impute the household income variable collected in the supplemental file *B* (the SHIW) to the matching records in the base file *A*, in order to obtain an original dataset *C* including all the variables of interest for the analysis. The inherent assumption in this procedure is that the random vector *Y* given *X* is independent of the random vector *Z* given *X*. The conditional independence assumption implies that *Y*'s relationship to *Z* can be totally inferred from *Y*'s relationship to *X* and *Z*'s relationship to *X*. Thus, the distributions of *X*, *Y*, and *Z* of the new file *C* must be identical to the distributions of *X*, *Y*, and *Z* empirically observed in the original files *A* and *B*. As a consequence, the best test to evaluate the quality of the statistical matching relies on the marginal distributions of the variables. As stated by Rässler (2002, 23), "A statistical match is said to be successful if the marginal and joint empirical distributions of *Z* and *Y* as they are observed in the donor samples are nearly the same in the statistically matched file". It should be clear, however, that "the statistical matching procedure does not generate new information about the conditional relationship of the *Y-Z* pair, but only reflects the assumptions used in creating the matched file" (Kadane 1978, 166).

The common variables $X = (X_1,...,X_P)$ shared by the original datasets are identified according to the following criteria: 1) they must have been classified and measured in the same (or very similar) way in both of the surveys. 2) They must have been observed for all the individuals included in the samples. 3) They can be assumed as possible determinants of health and social interaction in the base file. Based on suggestions from previous studies, we chose the following variables: gender, age, education, family size, number of children, region of residence, work status, sector of activity, and homeownership. The statistical matching was then performed through a regression imputation with random residuals. More specifically, the regression parameters of *Z* (i.e. the household income) on *X* were estimated on the SHIW. After this, a random residual was added to the regression prediction to obtain the imputed value of *z* for each $a = 1,...,n_A$ record in file *A*. Finally, the quality of the procedure was controlled by comparing, for each of the considered years, the conditional distribution of the household income given *X* in the new and the original files. The marginal distributions are not found to be statistically different[9].

Our final dataset *C* is a cross section sample of 50,618 observations. In this file, the level of household income "drawn" from the Survey on Household Income and Wealth carried out by the Bank of Italy is imputed to the $n_A$ statistical records included in the Istat Survey on Households.

---

[9] Distributions are available upon request to the authors.

# Appendix 2

| Table B1. Detailed description of variables | |
|---|---|
| *Dependent variable* | |
| Self-perceived good health | Individual assessment of health; 1 = good and very good |
| *Social capital: frequency of meeting with friends* | |
| Meetings with friends | 1 = every day or more times a week |
| *Instrumental variable* | |
| Mass attendance | Habit of attending church, 1 = once a week |
| Meetings with friends at the community level (average) | The mean value of the individual frequency of meetings with friends for each of the 6 categories of municipality size in each of the 20 Italian regions |
| *Demographic and socio-economic characteristics* | |
| Female | Gender of the respondent, 1 = female. **Reference group: male** |
| Single | Marital status of the respondent, 1= single. **Reference group: married** |
| Divorced | Marital status of the respondent, 1= divorced |
| Widowed | Marital status of the respondent, 1= widowed |
| Age31-40 | Age of the respondent, 1 = age between 31 and 40. **Reference group: age18-30** |
| Age41-50 | Age of the respondent, 1 = age between 41 and 50 |
| Age51-65 | Age of the respondent, 1 = age between 51 and 65 |
| Age>65 | Age of the respondent, 1 = age above 65 |
| Household size | Number of people who live in the family |
| Children0_5 | Number of children, 1 = if the number of children is aged between 0 and 5 years. **Reference group: no children**. |
| Children6_12 | Number of children, 1 = if the number of children is aged between 6 and 12 years |
| Children13_17 | Number of children, 1 = if the number of children is aged between 13 and 17 years |
| Elementary | Education of the respondent, 1 = completed elementary school (5 years). **Reference group: no education** |
| Junior high school | Education of the respondent, 1 = completed junior high school (8 years) |
| High school (diploma) | Education of the respondent, 1 = completed high school (13 years) |
| Bachelor's degree | Education of the respondent, 1 = university degree and/or doctorate (18 years and more) |
| Household income (ln) | Natural logarithm of imputed household income (sum of labour income, capital income and pensions) |
| Self-employed | Employment status of the respondent, 1 = self-employed. **Reference group: employed** |
| Unemployed | Employment status of the respondent, 1 = unemployed |
| Student | Employment status of the respondent, 1 = student |
| Retired | Employment status of the respondent, 1 = retired |
| Newspapers | Whether the respondent reads newspapers every day, 1 = yes |
| Homeowner | Whether the respondent owns a home outright, yes = 1 |
| Civil house | Whether the respondent lives in a council house, yes = 1 |
| Micro-criminality | Whether the respondent has even been pickpocketed, yes = 1 |
| *Size of municipality* | |
| Metropolis | Whether the respondent declares that he lives in a metropolitan area, yes=1. **Reference group: <2000** |
| Neighboring metropolis | Whether the respondent declares that he lives in a municipality neighbouring a metropolitan area, yes=1 |
| 2,000-10,000 | Whether the respondent declares that he lives in a municipality with 2,000-10,000 inhabitants, yes=1 |
| 10,000-50,000 | Whether the respondent declares that he lives in a municipality with 10,000-50,000 inhabitants, yes=1 |
| >50,000 | Whether the respondent declares that he lives in a municipality with more than 50,000 inhabitants, yes=1 |

Appendix 3

Table C1. Probit estimates of consumption of alcohol and water

|  | I | | II | | III | |
|---|---|---|---|---|---|---|
|  | Marg. Eff. | Std. Err. | Marg. Eff. | Std. Err. | Marg. Eff. | Std. Err. |
| Mass attendance | -0.004 | 0.003 | -0.002 | 0.001 | -0.001 | 0.005 |
| Meetings with friends | 0.009*** | 0.003 | 0.001 | 0.001 | 0.008* | 0.005 |
| Female | -0.001 | 0.003 | -0.001 | 0.001 | -0.005 | 0.004 |
| Single | -0.002 | 0.004 | -0.002 | 0.002 | -0.013* | 0.007 |
| Divorced | -0.006 | 0.007 | 0.000 | 0.003 | -0.009 | 0.011 |
| Widowed | -0.007 | 0.005 | -0.001 | 0.002 | -0.003 | 0.009 |
| Age31-40 | 0.006 | 0.005 | 0.002 | 0.002 | -0.018** | 0.008 |
| Age41-50 | -0.006 | 0.005 | -0.002 | 0.002 | -0.020** | 0.008 |
| Age51-65 | -0.001 | 0.006 | -0.000 | 0.002 | -0.026*** | 0.010 |
| Age > 65 | -0.007 | 0.007 | 0.003 | 0.003 | -0.029** | 0.012 |
| Household size | -0.003* | 0.001 | 0.000 | 0.000 | 0.006** | 0.002 |
| Children 0-5 | -0.007* | 0.004 | -0.003 | 0.002 | -0.012* | 0.006 |
| Children 6-12 | 0.002 | 0.003 | -0.003** | 0.001 | -0.009* | 0.005 |
| Children 13-17 | -0.003 | 0.004 | -0.001 | 0.001 | 0.001 | 0.006 |
| Elementary | -0.010* | 0.005 | -0.004* | 0.002 | -0.000 | 0.010 |
| Junior high school | -0.012* | 0.006 | -0.002 | 0.002 | 0.010 | 0.011 |
| High school (diploma) | -0.021*** | 0.007 | -0.003 | 0.003 | 0.009 | 0.012 |
| Bachelor's degree and beyond | -0.026*** | 0.007 | -0.004 | 0.003 | 0.009 | 0.015 |
| Household income (ln) | 0.007 | 0.006 | 0.001 | 0.002 | -0.003 | 0.009 |
| Self-employed | -0.001 | 0.004 | -0.003* | 0.002 | -0.007 | 0.007 |
| Unemployed | 0.001 | 0.006 | -0.003 | 0.002 | -0.041*** | 0.010 |
| Retired | 0.011*** | 0.004 | 0.003* | 0.002 | -0.006 | 0.007 |
| Student | -0.010 | 0.006 | 0.002 | 0.002 | -0.016 | 0.010 |
| Newspaper reader | -0.002 | 0.003 | 0.001 | 0.001 | 0.014*** | 0.005 |
| Homeowner | 0.007* | 0.004 | 0.000 | 0.001 | -0.001 | 0.006 |
| Civil house | -0.010*** | 0.003 | -0.002** | 0.001 | 0.022*** | 0.004 |
| Micro-criminality | -0.010 | 0.008 | -0.003 | 0.003 | 0.007 | 0.013 |
| *Size of municipality* | | | | | | |
| Metropolis | 0.010* | 0.006 | 0.000 | 0.002 | 0.005 | 0.009 |
| Neighbouring metropolis | 0.016** | 0.007 | -0.001 | 0.002 | 0.006 | 0.010 |
| 2000-10000 | 0.007 | 0.005 | 0.001 | 0.002 | 0.006 | 0.008 |
| 10000-50000 | 0.008 | 0.005 | -0.000 | 0.002 | -0.002 | 0.008 |
| >50000 | 0.003 | 0.006 | 0.002 | 0.002 | 0.009 | 0.009 |
| Regional dummies | Yes | | Yes | | Yes | |
| Pseudo R-squared | 0.01 | | 0.01 | | 0.01 | |
| Log-likelihood | -12694.89 | | -3106.95 | | -25963.06 | |
| No. of observations | 44204 | | 44294 | | 44366 | |

Note: dependent variable is *consumption of wine or beer with meals* a binary variable (1 = more than half a liter per day, 0 otherwise). In model II the dependent variable is *consumption of wine or alcohol outside meals* (1 = every day, 0 otherwise). In model III the dependent variable is *consumption of water* (1= more a liter per day, 0 otherwise). See Appendix 2 Table 1B for a detailed description of regressors. Regional dummies are omitted for space reasons. Standard errors are corrected for heteroskedasticity. The symbols ***, **, * denote that the coefficient is statistically different from zero at 1, 5 and 10 percent.

| Table C2 Ordered Probit estimates of friends' relationships (I), leisure (II) and economic situation (III) satisfactions | | | | | | |
|---|---|---|---|---|---|---|
| | I | | II | | III | |
| | Coeff. | Std. Err | Coeff. | Std. Err. | Coeff. | Std. Err. |
| Mass attendance | 0.067*** | 0.013 | 0.041*** | 0.012 | 0.066*** | 0.013 |
| Meetings with friends | 0.499*** | 0.013 | 0.270*** | 0.012 | 0.136*** | 0.013 |
| Female | 0.030*** | 0.012 | 0.020* | 0.011 | 0.008 | 0.012 |
| Single | -0.245*** | 0.018 | -0.023 | 0.018 | -0.227*** | 0.018 |
| Divorced | -0.133*** | 0.031 | -0.022 | 0.029 | -0.315*** | 0.030 |
| Widowed | -0.083*** | 0.022 | -0.012 | 0.022 | -0.095*** | 0.022 |
| Age31-40 | -0.123*** | 0.021 | -0.031 | 0.020 | -0.113*** | 0.021 |
| Age41-50 | -0.155*** | 0.023 | 0.025 | 0.022 | -0.161*** | 0.024 |
| Age51-65 | -0.152*** | 0.026 | 0.110*** | 0.025 | -0.229*** | 0.026 |
| Age > 65 | -0.285*** | 0.032 | 0.200*** | 0.031 | -0.255*** | 0.032 |
| Household size | -0.011* | 0.006 | 0.012** | 0.006 | -0.109*** | 0.006 |
| Children 0-5 | -0.050*** | 0.016 | -0.167*** | 0.015 | 0.023 | 0.017 |
| Children 6-12 | -0.020 | 0.013 | -0.056*** | 0.013 | 0.009 | 0.014 |
| Children 13-17 | 0.116*** | 0.015 | 0.078*** | 0.014 | 0.071*** | 0.015 |
| Elementary | 0.089*** | 0.025 | 0.106*** | 0.024 | 0.042* | 0.024 |
| Junior high school | 0.137*** | 0.029 | 0.173*** | 0.028 | 0.042 | 0.028 |
| High school (diploma) | 0.169*** | 0.031 | 0.238*** | 0.030 | 0.172*** | 0.031 |
| Bachelor's degree and beyond | 0.235*** | 0.038 | 0.343*** | 0.037 | 0.332*** | 0.038 |
| Household income (ln) | 0.025 | 0.024 | -0.190*** | 0.023 | 0.432*** | 0.024 |
| Self-employed | 0.045** | 0.018 | -0.130*** | 0.017 | 0.098*** | 0.018 |
| Unemployed | -0.056** | 0.026 | 0.031 | 0.025 | -0.445*** | 0.027 |
| Retired | 0.003 | 0.018 | 0.098*** | 0.018 | 0.088*** | 0.018 |
| Student | 0.025 | 0.028 | 0.064** | 0.027 | 0.012 | 0.028 |
| Newspaper reader | 0.144*** | 0.014 | 0.085*** | 0.013 | 0.131*** | 0.014 |
| Homeowner | 0.036** | 0.016 | 0.110*** | 0.015 | 0.133*** | 0.015 |
| Civil house | 0.018 | 0.011 | -0.003 | 0.011 | 0.039*** | 0.011 |
| Micro-criminality | -0.010 | 0.033 | -0.016 | 0.032 | -0.051 | 0.033 |
| *Size of municipality* | | | | | | |
| Metropolis | 0.049** | 0.023 | 0.017 | 0.022 | 0.017 | 0.022 |
| Neighbouring metropolis | 0.017 | 0.027 | -0.024 | 0.026 | 0.024 | 0.027 |
| 2000-10000 | 0.027 | 0.021 | 0.024 | 0.020 | 0.018 | 0.021 |
| 10000-50000 | 0.033 | 0.022 | 0.004 | 0.021 | 0.020 | 0.021 |
| >50000 | 0.014 | 0.023 | -0.000 | 0.022 | -0.002 | 0.023 |
| Regional dummies | Yes | | Yes | | Yes | |
| Pseudo R-squared | 0.04 | | 0.02 | | 0.05 | |
| Log-likelihood | -42031.73 | | -49655.62 | | -42506.68 | |
| No. of observations | 44196 | | 44120 | | 44278 | |

Note: we regard a question on a four-point scale: "Consider the last twelve months. Are you satisfied with the following domains of your life?". The following areas of life are considered: friends' relationships, leisure and economic situation. The responses are: "Very satisfied", "Quite satisfied", "Not very satisfied", "Not at all satisfied". We recode the answer on a scale from 1 to 4, with 1 being "Not at all satisfied" and 4 being "Very satisfied". The symbols ***, **, * denote that the coefficient is statistically different from zero at 1, 5 and 10 percent.

# Appendix 4

Table D1. First stage instrumental variables estimates (dependent variable: meetings with friends)

|  | I | | II | | III | |
|---|---|---|---|---|---|---|
|  | IV Probit | | Biprobit | | 2SLS | |
|  | Coeff. | Std. Err. | Coeff. | Std. Err | Coeff. | Std. Err. |
| Mass attendance | 0.031*** | 0.005 | 0.104*** | 0.016 | 0.031*** | 0.004 |
| Meetings with friends at community level | 0.933*** | 0.046 | 2.908*** | 0.150 | 0.933*** | 0.045 |
| Female | -0.066*** | 0.004 | -0.227*** | 0.014 | -0.066*** | 0.004 |
| Single | 0.092*** | 0.007 | 0.357*** | 0.024 | 0.092*** | 0.007 |
| Divorced | -0.003 | 0.011 | 0.000 | 0.034 | -0.003 | 0.011 |
| Widowed | -0.021** | 0.008 | -0.018 | 0.025 | -0.021** | 0.010 |
| Age31-40 | -0.082*** | 0.008 | -0.359*** | 0.029 | -0.082*** | 0.007 |
| Age41-50 | 0.032*** | 0.012 | -0.006 | 0.043 | 0.032*** | 0.012 |
| Age51-65 | -0.013 | 0.013 | -0.140*** | 0.045 | -0.013 | 0.013 |
| Age > 65 | -0.087*** | 0.015 | -0.338*** | 0.050 | -0.087*** | 0.015 |
| Household size | -0.007*** | 0.002 | -0.014* | 0.008 | -0.007*** | 0.002 |
| Children 0-5 | -0.064*** | 0.006 | -0.218*** | 0.019 | -0.064*** | 0.006 |
| Children 6-12 | -0.011** | 0.005 | -0.046*** | 0.016 | -0.011** | 0.005 |
| Children 13-17 | 0.005 | 0.005 | 0.009 | 0.018 | 0.005 | 0.005 |
| Elementary | 0.056*** | 0.009 | 0.162*** | 0.028 | 0.56*** | 0.010 |
| Junior high school | 0.075*** | 0.010 | 0.215*** | 0.033 | 0.075*** | 0.012 |
| High school (diploma) | 0.058*** | 0.012 | 0.165*** | 0.037 | 0.058*** | 0.012 |
| Bachelor's degree and beyond | 0.062*** | 0.014 | 0.174*** | 0.047 | 0.062*** | 0.015 |
| Household income (ln) | -0.042*** | 0.009 | -0.131*** | 0.030 | -0.042*** | 0.008 |
| Self-employed | 0.015** | 0.007 | 0.044** | 0.022 | 0.015** | 0.006 |
| Unemployed | 0.013 | 0.010 | 0.082** | 0.037 | 0.013* | 0.008 |
| Retired | 0.061*** | 0.007 | 0.161*** | 0.021 | 0.061*** | 0.007 |
| Student | 0.051*** | 0.010 | 0.429*** | 0.052 | 0.051*** | 0.006 |
| Newspaper reader | 0.051*** | 0.005 | 0.168*** | 0.017 | 0.051*** | 0.005 |
| Homeowner | 0.038*** | 0.006 | 0.128*** | 0.019 | 0.038*** | 0.006 |
| Civil house | 0.002 | 0.004 | 0.004 | 0.014 | 0.002 | 0.004 |
| Micro-criminality | 0.012 | 0.012 | 0.043 | 0.041 | 0.012 | 0.012 |
| *Size of municipality* | | | | | | |
| Metropolis | -0.000 | 0.008 | -0.004 | 0.028 | -0.000 | 0.008 |
| Neighbouring metropolis | -0.007 | 0.010 | -0.013 | 0.033 | -0.007 | 0.010 |
| 2000-10000 | -0.005 | 0.008 | -0.023 | 0.026 | -0.005 | 0.008 |
| 10000-50000 | -0.003 | 0.008 | -0.014 | 0.027 | -0.003 | 0.008 |
| >50000 | -0.003 | 0.008 | -0.011 | 0.028 | -0.003 | 0.008 |
| Regional dummies | Yes | | Yes | | Yes | |
| No. of observations | 44704 | | 44704 | | 44704 | |

Note. The symbols ***, **, * denote that the coefficient is statistically different from zero at 1, 5 and 10 percent.